\documentclass[12pt]{article}
\usepackage{amsmath, amsthm, amssymb, amsfonts, enumerate,bm,url}
\usepackage[figuresright]{rotating}
\usepackage{lscape}
\usepackage{multirow}
\usepackage{natbib}

\begin{document}
\newcommand{\abs}[1]{\left\vert#1\right\vert}
\newcommand{\set}[1]{\left\{#1\right\}}
\newcommand{\eps}{\varepsilon}
\newcommand{\To}{\rightarrow}
\newcommand{\inv}{^{-1}}
\newcommand{\ihat}{\hat{\imath}}
\newcommand{\var}{\mbox{Var}}
\newcommand{\sd}{\mbox{SD}}
\newcommand{\cov}{\mbox{Cov}}
\newcommand{\f}{\frac}
\newcommand{\fI}[1]{\frac{1}{#1}}
\newcommand{\what}[1]{\widehat{#1}}
\newcommand{\hhat}[1]{\what{\what{#1}}}
\newcommand{\wtilde}[1]{\widetilde{#1}}
\newcommand{\bdot}{\bm{\cdot}}
\newcommand{\Th}{\theta}
\newcommand{\qmq}[1]{\quad\mbox{#1}\quad}
\newcommand{\qm}[1]{\quad\mbox{#1}}
\newcommand{\mq}[1]{\mbox{#1}\quad}
\newcommand{\tr}{\mbox{tr}}
\newcommand{\logit}{\mbox{logit}}
\newcommand{\noi}{\noindent}
\newcommand{\bni}{\bigskip\noindent}
\newcommand{\bul}{$\bullet$ }
\newcommand{\bias}{\mbox{bias}}
\newcommand{\conv}{\mbox{conv}}
\newcommand{\spn}{\mbox{span}}
\newcommand{\colspace}{\mbox{colspace}}
\newcommand{\mA}{\mathcal{A}}
\newcommand{\mF}{\mathcal{F}}
\newcommand{\mH}{\mathcal{H}}
\newcommand{\mI}{\mathcal{I}}
\newcommand{\mJ}{\mathcal{J}}
\newcommand{\mN}{\mathcal{N}}
\newcommand{\mR}{\mathcal{R}}
\newcommand{\mT}{\mathcal{T}}
\newcommand{\mX}{\mathcal{X}}
\newcommand{\mS}{\mathcal{S}}
\newcommand{\bbR}{\mathbb{R}}
\newcommand{\fweI}{FWE$_I$}
\newcommand{\fweII}{FWE$_{II}$}
\newcommand{\fdp}{\mbox{FDP}}
\newcommand{\vphi}{\varphi}
\newcommand{\oN}{\overline{N}}

\newtheorem{theorem}{Theorem}[section]
\newtheorem{corollary}{Corollary}[section]
\newtheorem{conjecture}{Conjecture}[section]
\newtheorem{proposition}{Proposition}[section]
\newtheorem{lemma}{Lemma}[section]
\newtheorem{definition}{Definition}[section]
\newtheorem{example}{Example}[section]
\newtheorem{remark}{Remark}[section]

\title{{\bf\Large A New Approach to Designing Phase I-II Cancer Trials for Cytotoxic Chemotherapies}}

\author{Jay Bartroff\footnote{Department of Mathematics, University of Southern
           California, 3620 South Vermont Avenue, KAP 108, Los
           Angeles, CA 90089, U.S.A., \textit{email:} bartroff@usc.edu}, Tze Leung Lai\footnote{Department of Statistics, Sequoia Hall, Stanford
            University, Stanford, CA 94305, U.S.A., \textit{email:} lait@stanford.edu}, and Balasubramanian Narasimhan\footnote{Department of Statistics, Sequoia Hall, Stanford
            University, Stanford, CA 94305, U.S.A., \textit{email:} naras@stat.stanford.edu}}

\date{}
\maketitle

\begin{abstract}
Recently there has been much work on early phase cancer designs that incorporate both toxicity and efficacy data, called Phase~I-II designs because they combine elements of both phases.  However, they do not explicitly address the Phase~II hypothesis test of $H_0: p\le p_0$, where $p$ is the probability of efficacy at the estimated maximum tolerated dose (MTD) $\what{\eta}$ from Phase~I and $p_0$ is the baseline efficacy rate. Standard practice for Phase~II remains to treat $p$ as a fixed, unknown parameter and to use Simon's 2-stage design with all patients dosed at $\what{\eta}$.   We propose a Phase~I-II design that addresses the uncertainty in the estimate $p=p(\what{\eta})$ in $H_0$ by using sequential generalized likelihood theory. Combining this with a Phase~I design that incorporates efficacy data, the Phase~I-II design provides a common framework that  can be used all the way from the first dose of Phase~I through the final accept/reject decision about $H_0$ at the end of Phase~II, utilizing both toxicity and efficacy data throughout.  Efficient group sequential testing is used in Phase~II that allows for early stopping to show treatment effect or futility. The proposed Phase~I-II design thus removes the artificial barrier between Phase~I and Phase~II, and fulfills the objectives of searching for the MTD and testing if the treatment has an acceptable response rate to enter into a Phase~III trial.
\end{abstract}

%\keywords{cancer trials, generalized likelihood ratio, group sequential, isotonic regression, maximum tolerated dose, Phase~I, Phase~II}

%\footnotetext[1]{Department of Mathematics, University of Southern
%           California, 3620 South Vermont Avenue, KAP 108, Los
%           Angeles, CA 90089, U.S.A., \textit{email:} bartroff@usc.edu}
%           
%\footnotetext[2]{Department of Statistics, Sequoia Hall, Stanford
%            University, Stanford, CA 94305, U.S.A.}
%\footnotetext[3]{\textit{email:} lait@stanford.edu}
%\footnotetext[4]{\textit{email:} naras@stat.stanford.edu}

\section{Introduction}\label{sec:intro}

In typical Phase~I studies in the development of relatively benign drugs, the drug is initiated at low doses and subsequently escalated to show safety at a level where some positive response occurs, and healthy
volunteers are used as study subjects. This paradigm does not work for
diseases like cancer, for which a non-negligible probability of severe
toxic reaction has to be accepted to give the patient some chance of a
favorable response to the treatment. Therefore patients (rather than healthy volunteers) are used as study subjects, and it is widely accepted that some degree of
toxicity must be tolerated to experience any substantial therapeutic
effects. Hence, an acceptable proportion $q$ of patients experiencing \textit{dose limiting toxicities} (DLTs) is generally agreed on before the trial, which depends on
the type and severity of the DLT; the dose resulting in this
proportion is thus referred to as the \textit{maximum tolerated dose} (MTD). In addition to the explicitly stated objective of determining the MTD, a Phase~I cancer trial also has the implicit goal of safe treatment of the patients in the trial. However, the aims of treating patients in the trial and generating an efficient design to estimate the MTD for future patients often run counter to each other. Commonly used designs in Phase~I cancer trials implicitly place their focus on the safety of the patients in the trial, beginning from a conservatively low starting dose and  escalating cautiously.

\citet{Bartroff10b,Bartroff11d} have given a review of model-based methods to design Phase~I cancer trials and proposed a general framework that incorporates both ``individual'' and ``collective'' ethics into the design of the trial. We have also developed a new design which minimizes a risk function composed of two terms, with one representing the individual risk of the current dose and the other representing the collective risk, and have shown that it performs better than existing model-based designs in accuracy of the MTD estimate at the end of the trial, and toxicity and overdose rates of patients in the trial, and loss functions reflecting the individual and collective ethics.

The MTD determined from a Phase~I study is used in a subsequent Phase~II study, in which ``a cohort of patients is treated,
and the outcomes are related to the prespecified target or bar. If the
results meet or exceed the target, the treatment is declared worthy of
further study; otherwise, further development is stopped. This has
been referred to as the `go/ no go' decision''
\citep[][p.~927]{Vickers07}. The most widely used designs for these single-arm Phase~II trials are Simon's two-stage design \citep{Simon89}, which allows early stopping of the trial if the treatment has not shown
beneficial effect, that is measured by a Bernoulli proportion. Simon considered the design that stops for futility (i.e., accepts the null hypothesis~$H_0$ in \eqref{SimonHyp})
 after $n_1$ patients if the number of patients exhibiting positive
treatment effect is $r_1 (\le n_1)$ or fewer, and otherwise treats an
additional $n_2$ patients and rejects the treatment (again, accepts $H_0$) if and only if the
number of patients exhibiting positive treatment effect is $r (\le
n_1+n_2)$ or fewer. Simon's design requires that a null proportion $p_0$,
representing some ``uninteresting'' level of positive treatment
effect, and an alternative $p_1>p_0$ be specified.  The null
hypothesis is 
\begin{equation}\label{SimonHyp}
H_0:p\le p_0,\end{equation} where $p$ denotes the probability of
positive treatment effect. The type~I and II error probabilities
$\alpha=P_{p_0}(\mbox{Reject $H_0$})$, $\beta=P_{p_1}(\mbox{Accept
  $H_0$})$ and the expected sample size $E_{p_0}N$ can be
computed for any design of this form, which can be
represented by the parameter vector $(n_1,n_2,r_1,r)$. Using computer
search over these integer-valued parameters, \citet{Simon89} tabulated the
optimal designs in his Tables~1 and 2 for different values of
$(p_0,p_1)$. Simon's design has been generalized by \citet{Jung01,Jung04} who also give a graphical method of selecting from among the admissible designs, Simon's original procedure being one of them, and by \citet{Lu05} to allow for partial responses. Whether the new treatment is declared promising in a Phase~II trial
depends strongly on the prescribed $p_0$ and $p_1$. The sample size~$m$ of a typical Phase~I trial and the maximum sample size $M=n_1+n_2$ of a typical Phase~II trial are relatively small, 20-30 for Phase~I and no more than 60 for Phase~II. \citet{Vickers07} conclude that  uncertainty in the choice of $p_0$ and
$p_1$ can increase the likelihood that (a) a treatment with no viable
positive treatment effect proceeds to Phase~III, or (b) a treatment
with positive treatment effect is abandoned at Phase~II.

\subsection{An integrated approach to dose finding and testing for efficacy}\label{sec:integrated}
In Sections~\ref{sec:new-approach} and \ref{sec:ordered} we address these issues concerning the design of early-phase single-arm cancer trials by developing a novel seamless Phase~I-II trial design that uses efficient statistical methods for the design and analysis of the integrated trial, subject to ethical and sample size constraints. The data from the trial are toxicity and efficacy outcomes at various doses and consist of $(x_i,y_i,z_i)$, $i=1,\ldots,N$, where $N$ is the Phase~I-II total sample size, $x_i$ denotes the dose given to the $i$th subject, $y_i=1$ or $0$ according to whether a DLT occurs or not, and $z_i=1$ or $0$ according to whether the subject responds to the treatment. For cytotoxic treatments, both the dose-toxicity curve $P(y_i=1|x_i=x)$ and the dose-response curve $P(z_i=1|x_i=x)$ increase with the dose~$x$, and therefore the MTD is the most efficacious dose subject to a prespecified probability~$q$ of severe toxic reaction. Whereas the objective of a traditional Phase~I cancer trial is to estimate the MTD, denoted by $\eta$, from $(x_i,y_i)$, $i=1,
\ldots,m$,  and that of the ensuing Phase~II trial with maximum sample size $M$ is to test if the response rate exceeds some prespecified level~$p_0$ when all patients in the trial are assigned dose $\what{\eta}$, which is the MTD estimate from the Phase~I trial, our integrated design continues sequential estimation of $\eta$ throughout the trial with total maximum sample size $m+M$ and uses an efficient  group sequential test of the null hypothesis that the response rate at $\eta$ does not exceed $p_0$. In Section~\ref{sec:new-approach} we consider commonly used logisitic regression models for dose-toxicity and dose-response relationships to pinpoint the basic ideas. Section~\ref{sec:ordered} removes the parametric assumptions and extends the methodology to dose-toxicity and dose-response relationships that are only assumed to be monotone. Simulation studies in Section~\ref{sec:sim} demonstrate the advantages of the integrated design, and Section~\ref{sec:imp+theory} describes the underlying theory and implementation details.

\subsection{Review of current methods using toxicity and efficacy/response data}

\citet{Gooley94} suggested using efficacy and toxicity data together, and performed simulations to compare the operating characteristics of three ad-hoc designs. \citet{Thall98} proposed a design combining binary toxicity data $y_i$ and trinomial response data $z_i=0,1$, or $2$ for no, moderate, or severe response, respectively, into a single trinomial variable 
\begin{equation}\label{trinom}
 w_i=\begin{cases}
0,&\mbox{if $z_i=0$ and $y_i=0$} \\
1,&\mbox{if $z_i=1$ and $y_i=0$} \\
2,&\mbox{if $z_i=2$ or $y_i=1$.}
\end{cases}
\end{equation} Using a proportional odds regression model for $w_i$ on dose $x_i$ with a prior distribution on its unknown parameters, a Bayesian posterior calculation along the lines of O'Quigley, Pepe, and Fisher's~\citeyearpar{OQuigley90} continual reassessment method (CRM) is performed to calculate the acceptability of the available discrete dose levels and escalate or de-escalate the current dose level. For a similar setting, \citet{OQuigley01} proposed a Phase~I design for HIV trials in which binary efficacy $z_i$ and toxicity $y_i$ variables are combined into a single trinomial variable~\eqref{trinom} in which we now set $w_i=2$ if $y_i=1$. A CRM-like calculation is used to treat the current patient at the posterior estimate of the dose maximizing the probability of simultaneous efficacy and non-toxicity.

For efficacy and toxicity measurements, \citet{Ivanova03} proposed an up-and-down design which assigns doses in pairs on a discrete set of dose levels. \citet{Braun02} proposed a bivariate version of CRM in which a bivariate joint distribution is chosen for $(y_i,z_i)$, and the target dose is defined to be the one minimizing the expected Euclidean
distance to pre-specified toxicity and efficacy rates, with respect to a chosen noninformative posterior distribution.  In particular, the bivariate distribution of  \citet{Arnold91} which gives Bernouilli conditional distributions of $y_i$ given $z_i$, and vice-versa, was recommended. \citet{Thall04} proposed a different method for combining efficacy and toxicity responses. First, marginal efficacy and toxicity
curves are assumed which are then combined using a Gaussian or Gumbel copula; this approach differs from Braun's method that specifies the conditional distributions rather than the marginals. Doses are then selected using  ``trade-off contours'' in the two-dimensional space of outcome probabilities on which the outcomes are equally desirable.  \citet{Thall08} extend this method to allow for the inclusion of patient-specific covariates.

Even when the designs summarized above are called ``Phase~I-II'' designs, it is because they incorporate efficacy (or tumor response) data. They do not address testing the efficacy hypothesis that is the purpose of typical Phase II cancer studies, for which the standard practice is to use Simon's 2-stage design following the dose-finding portion. Moreover, this skirts the issue of uncertainty in the estimated MTD used in the null hypothesis in Phase~II, as well as ignores toxicity outcomes that are available during Phase~II which could help improve this estimate, especially since the Phase~I sample size is usually small. The innovative Phase~I-II design proposed herein aims at rectifying these issues, and hence provides a common framework that  can be used all the way from Phase~I through the final accept/reject decision about the null hypothesis on efficacy in the Phase~II portion of the study, utilizing both toxicity and efficacy data for dose finding while performing efficient group sequential testing of the null hypothesis.  

\section{An integrated approach to designing early-phase cancer clinical trial designs}\label{sec:new-approach}

A widely-used model for the dose-toxicity curve is the logistic regression model
\begin{equation}\label{logit_tox}
P(y_i=1|x_i=x)=F(x;\bm{\theta}):=1/\left(1+e^{-(\theta_1+\theta_2 x)}\right),
\end{equation} where $\bm{\theta}=(\theta_1,\theta_2)$ and it is assumed that $\theta_2>0$ (i.e., probability of toxicity increases with dose), for which the MTD is given by $\eta=[\log(q/(1-q))-\theta_1]/\theta_2$. Under (\ref{logit_tox}), the estimate $\what{\eta}$ based on $(x_i,y_i)$, $i=1,
\ldots,m$, can be obtained by maximum likelihood, which is equivalent to logistic regression. Similarly, we can model the dose-response curve by 
\begin{equation}\label{logit_eff}
P(z_i=1|x_i=x)=p(x;\bm{\psi}):=1/\left(1+e^{-(\psi_1+\psi_2 x)}\right),
\end{equation} under which the probability~$p$ of the response in the null hypothesis $H_0:p\le p_0$ of the traditional Phase~II cancer trial is actually $p(\what{\eta};\bm{\psi})$. The difference between $\what{\eta}$ and $\eta$ is completely ignored in currently used designs, and the toxicity outcomes in the Phase~II trial are also ignored. Combining the toxicity outcomes in Phase~II with those in Phase~I can improve the estimate of $\eta$, especially since the Phase~I sample size is small.  Changing the null hypothesis to 
\begin{equation}\label{realH0}
H_0: p(\eta;\bm{\psi})\le p_0\end{equation} not only takes into consideration the uncertainties in $\what{\eta}$ as an estimate of $\eta$ but also leads to continual updating of $\eta$ with toxicity outcomes in the Phase~II trial if one uses a generalized likelihood ratio (GLR) test. Moreover, the GLR test also uses the Phase~I efficacy outcomes $z_i$, $i=1,\ldots,m$.

\subsection{The first phase of the Phase~I-II trial}
The first phase of the new Phase~I-II (or dose-finding) design involves only the dose-toxicity data, but not the responses $z_i$. We can use traditional methods or recent advances in Phase~I cancer trial designs to perform dose escalation; see Section~\ref{sec:imp} and the references therein for details. At the end of the Phase~I trial, we compute the maximum likelihood or Bayes estimates $\wtilde{\bm{\theta}}$, $\wtilde{\eta}$, and $\wtilde{\bm{\psi}}$ of $\bm{\theta}$, $\eta$, and $\bm{\psi}$. Let $\mF_0$ denote the Phase~I data $(x_1,y_1,z_1), \ldots, (x_{m},y_{m},z_{m})$.

\subsection{The ensuing group sequential design to test efficacy and re-estimate $\eta$} \label{sec:grp-sec-des}
After this initial group of $m$ patients, the proposed design switches to a group sequential scheme, with specified group sizes $m_1,\ldots,m_K$ (e.g., $m_1=\ldots=m_K$ gives constant group size sampling). The group sequential scheme updates the MTD estimate via MLE at the $k$th interim analysis with an additional batch of size $m_k$ of dose-toxicity data $(x_{\tau_{k-1}+1},y_{\tau_{k-1}+1}),\ldots (x_{\tau_{k}},y_{\tau_{k}})$, where\begin{equation}\label{tau_k}
\tau_k=m+\sum_{i=1}^k m_i,\quad k=1,\ldots,K.
\end{equation}
It also uses all the observed data $(x_i,y_i,z_i)$, $1\le i\le \tau_k$, to perform a group sequential GLR test of $H_0:p(\eta;\bm{\psi})\le p_0$ at the $k$th interim analysis, where $p(x;\bm{\psi})$ is defined by (\ref{logit_eff}).  \citet{Lai04} have developed a methodology of nearly optimal group sequential tests, which use versatile and asymptotically efficient GLR test statistics and stopping boundaries. In conjunction with GLR statistics, maximum likelihood (rather than Bayes) estimates of $\eta$ are used for sequential updating of the estimated MTD.

To simplify the description, we begin by assuming that $y_i$ and $z_i$ are independent; this assumption will be removed in Section~\ref{sec:dependent}. Let $\ell_k(\bm{\psi})$ denote the log-likelihood function for $\bm{\psi}$ at the $k$th interim analysis, which because of the independence assumption only depends on the $z_i$ and not the $y_i$:
$$\ell_k(\bm{\psi})=\log\left\{\prod_{i=1}^{\tau_k}  p(x_i;\bm{\psi})^{z_i} [1-p(x_i;\bm{\psi})]^{1-z_i}\right\}=\sum_{i=1}^{\tau_k} \left\{z_i(\psi_1+\psi_2 x_i)-\log(1+e^{\psi_1+\psi_2 x_i}) \right\}.$$
Let $\what{\bm{\psi}}_{k}$ be an MLE maximizing this, $\what{\bm{\theta}}_k=(\what{\theta}_{k,1},\what{\theta}_{k,2})$ be an MLE of $\bm{\theta}$ based on the data up to and including the $k$th interim analysis, $\what{\eta}_{k}=(\logit(q)-\what{\theta}_{k,1})/\what{\theta}_{k,2}$, and
\begin{equation}
\mathcal{S}_k^j=\{\bm{\psi}: p(\what{\eta}_{k};\bm{\psi})=p_j\}\qm{for $0\le k\le K$, $j=0,1$,}
\end{equation} where $\what{\eta}_0=\wtilde{\eta}$, $p_1>p_0$ and $H_1: p(\eta;\bm{\psi})\ge p_1$ is the alternative hypothesis. The choice of $p_1$ will be discussed in Section~\ref{sec:imp}.

As will be explained in Section~\ref{sec:theory}, we can compute at the $k$th interim analysis the test statistics
\begin{equation}\label{GLRs}
\ell_{k,j}=\min_{\bm{\psi}\in\mathcal{S}_k^j}\left[ \ell_k(\what{\bm{\psi}}_{k})- \ell_k(\bm{\psi})\right],\quad j=0,1,
\end{equation} so that the group sequential test stops and rejects $H_0$ at interim analysis $k<K$ if
\begin{equation}\label{b-rej}
p(\what{\eta}_{k},\what{\bm{\psi}}_{k})>p_0\qmq{and} \ell_{k,0}\ge b,
\end{equation} and early stopping for futility (accepting $H_0$) at analysis $k<K$ can also occur if
\begin{equation}\label{btild-rej}
p(\what{\eta}_{k},\what{\bm{\psi}}_{k})<p_1\qmq{and} \ell_{k,1}\ge \wtilde{b}.
\end{equation} The test rejects $H_0$ at the $K$th analysis if 
\begin{equation}\label{c-rej}
p(\what{\eta}_{K},\what{\bm{\psi}}_{K})>p_0\qmq{and} \ell_{K,0}\ge c.
\end{equation} The thresholds $b,\wtilde{b}$, and $c$ are chosen so that  
\begin{equation}\label{stage2-typeI}
\max_{\bm{\psi}\in\mathcal{S}_0^0} P_{\wtilde{\bm{\theta}},\bm{\psi}}(\mbox{$H_0$ rejected}|\mathcal{F}_0)= \alpha
\end{equation}
and the power
\begin{equation}\label{stage2-pow}
\min_{\bm{\psi}\in\mathcal{S}_0^1} P_{\wtilde{\bm{\theta}},\bm{\psi}}(\mbox{$H_0$ rejected}|\mathcal{F}_0)
\end{equation} 
is close to $1-\beta$, as in \citet{Lai04}, \citet{Bartroff08}, and \citet{Bartroff08c}. Details and software for implementation are given in Section~\ref{sec:imp}.

\subsection{Modeling the dependence between $y_i$ and $z_i$}\label{sec:dependent}

We can model the dependence between $y_i$ and $z_i$ by replacing the marginal model \eqref{logit_eff} by the following model for the conditional distribution of $z_i$ given $y_i$:
\begin{align}
P(z_i=1|y_i=0, x_i)&=1/\left(1+e^{-(\psi_1^0+\psi_2^0 x_i)}\right) \label{z|y=0}\\
P(z_i=1|y_i=1, x_i)&=1/\left(1+e^{-(\psi_1^1+\psi_2^1 x_i)}\right)\label{z|y=1}
\end{align} with parameters $\bm{\psi}^0=(\psi_1^0,\psi_2^0)$ and $\bm{\psi}^1=(\psi_1^1,\psi_2^1)$. Generalizing \eqref{realH0} to include \eqref{z|y=0}-\eqref{z|y=1}, the null hypothesis is that the probability of efficacy at dose $x=\eta$ is less than or equal to $p_0$, i.e.,
\begin{equation}\label{H0.dependent}
H_0: \frac{1-q}{1+e^{-(\psi_1^0+\psi_2^0 \eta)}} + \frac{q}{1+e^{-(\psi_1^1+\psi_2^1 \eta)}}\le p_0,
\end{equation} noting that $F(\eta;\bm{\theta})=q$. This null hypothesis is an extension of that in Section~\ref{sec:grp-sec-des} and can again be tested by sequential GLR theory.

\subsection{Modifications for discrete dose levels}

In practice the dose levels in dose-finding studies of cancer drugs are usually chosen before the trial from a finite set 
\begin{equation}\label{levels}
\Lambda=\{\lambda_1,\ldots,\lambda_d\},\qmq{where}\lambda_1< \lambda_2<\ldots <\lambda_d,
\end{equation}
unlike the continuous doses we have assumed so far. In this case the MTD has to be redefined as 
\begin{equation}\label{discMTD}
\eta=\begin{cases}
\max\{\lambda\in \Lambda: F(\lambda;\bm{\theta})\le q\},&\mbox{if $F(\lambda_i;\bm{\theta})\le q$ for some $i$}\\
\lambda_1,&\mbox{otherwise.}
\end{cases}
\end{equation}
Putting this modified definition of $\eta$ in \eqref{realH0} or \eqref{H0.dependent}, we can still apply the group sequential GLR test of Section~\ref{sec:grp-sec-des} or \ref{sec:dependent}, in which we also modify the definition of $\what{\eta}_k$ accordingly to be $\Lambda$-restricted. That is, $\what{\eta}_k$ is the smallest $\lambda_j\in\Lambda$ maximizing the likelihood $\ell_k$ up through the $k$th interim analysis, and we set $x_i=\what{\eta}_k$ for $i=\tau_k+1, \ldots,\tau_{k+1}$. Note that the group sequential GLR test is based on all the observed data $(x_i,y_i,z_i)$ up to the time of an interim analysis, irrespective of how the $x_i$ are chosen and therefore no additional modifications are needed. 

Since $\Lambda$ is discrete, one can use more robust specification of the dose-toxicity and/or dose-response curve than the logisitic regression models \eqref{logit_tox} and \eqref{logit_eff}. Details are given in the next section. For samples of the size typically used in early-phase cancer trials, however, one usually does not have enough data to detect departures from these ``working models.'' In addition, the initial phase of a dose-finding study for cytotoxic chemotherapies is often very conservative, to avoid causing harm to patients before observing how the new treatment actually works in human subjects. This explains the popularity of the widely-used, although inefficient, 3+3 designs. A more efficient alternative is to use a 2-stage Phase~I design in which a more cautious design is used for the first stage before switching to a parametric model-based design in the second stage; see \cite[Section~4.2]{Bartroff10b}.  Once we have zoomed in on a range around the MTD that is narrow relative to the original dose range, the logistic model is actually quite robust because it can be viewed as a locally linear regression model around the MTD, adjusted with the logit link for Bernoulli outcomes. What this means is that one only needs to be  concerned with the choice of the design levels $x_i$ to ensure such robustness in the locally logit-linear model. Thus, the GLR test statistic can be restricted only to those $x_i$ that are within a certain distance from $\what{\eta}_k$ at the $k$th interim analysis.

\section{Extension to monotone dose-toxicity and dose-response relationships}\label{sec:ordered}

In many dose-finding trials, the number of discrete dose levels \eqref{levels} is relatively small. For this situation, in this section we develop an approach similar to the Bayesian models of  \citet{Yin06} and  \citet{Yin09b} where the probabilities of toxicity and efficacy are order-restricted, but in a frequentist setting.  Assume for now that $y_i$ and $z_i$ are independent; the general case will be covered below in Section~\ref{sec:disc.dep}. Because the number of dose levels is small, we also assume that all the levels have been used at least once during Phase~I; if this does not hold then only the used dose levels are carried forward into Phase~II. Instead of the parameterization  by the toxicity and efficacy parameters $\bm{\theta}$ and $\bm{\psi}$, we parameterize by the toxicity and efficacy probabilities
\begin{equation}\label{disc.nonparm}
\phi_i=P(y=1|x=\lambda_i),\quad \pi_i=P(z=1|x=\lambda_i),\quad i=1,\ldots,d.
\end{equation} The MTD \eqref{discMTD} can then be written
\begin{equation*}
\eta=\lambda_{i^*}\qmq{where} i^*=\begin{cases}
\max\{i: \phi_i\le q\},&\mbox{if $\phi_i\le q$ for some $i$}\\
1,&\mbox{otherwise}
\end{cases}
\end{equation*} 
so that the Phase~II null and alternative hypotheses can be expressed as
\begin{equation*}
H_0: \pi_{i^*}\le p_0\qmq{vs.} H_1: \pi_{i^*}\ge p_1.
\end{equation*} 

\subsection{Order-restricted MLE and GLR statistics}
Letting $x_t=\lambda_{i_t}$ denote the $t$-th dose, $t=1,\ldots,\tau_K$ with $\tau_k$ given by \eqref{tau_k}, and $\bm{\pi}=(\pi_1,\ldots,\pi_d)$, the log-likelihood at the $k$th interim analysis of Phase~II under the independence assumption is
\begin{equation}\label{discLL}
\ell_k(\bm{\pi})=\log\left\{\prod_{t=1}^{\tau_k}\pi_{i_t}^{z_t}(1-\pi_{i_t})^{1-z_t}\right\}.
\end{equation} The order-restricted MLE $\what{\bm{\pi}}_k=(\what{\pi}_{1,k},\ldots,\what{\pi}_{d,k})$ maximizing \eqref{discLL} subject to $\pi_1\le\ldots\le \pi_d$ is given by the formula 
\begin{equation}\label{ordMLE}
\what{\pi}_{i,k}=\min_{j'\ge i}\max_{j\le i}\left( \frac{S_{j,k}+\cdots+S_{j',k}}{\nu_{j,k}+\cdots+\nu_{j',k}}\right),\quad i=1,\ldots,d,
\end{equation} where $S_{i,k}=\sum_{t=1}^{\tau_k}z_t1\{i_t=i\}$ is the sum of the efficacy responses at level $i$ and $\nu_{i,k}=\sum_{t=1}^{\tau_k}1\{i_t=i\}$ is the number of patients that have been dosed at level $i$ up through the $k$th analysis \citep[][p.~52]{Silvapulle04}. An analogous formula holds for the order-restricted MLE of the toxicity probabilities $\what{\bm{\phi}}_k=(\what{\phi}_{1,k},\ldots,\what{\phi}_{d,k})$. These order-restricted MLEs can be computed by solving the minimization-maximization problem in \eqref{ordMLE} or, equivalently, by using the well known Pool Adjacent Violators Algorithm (PAVA); see \cite[Section~2.4]{Silvapulle04}. 

The order-restricted MLE of the MTD at the $k$th interim analysis can be defined as 
\begin{equation}\label{iso.MLE}
\what{\eta}_k=\lambda_{\what{i}^*_k},  \qmq{where} \what{i}^*_k=\begin{cases}
\max\{i: \what{\phi}_{i,k}\le q\},&\mbox{if $\what{\phi}_{i,k}\le q$ for some $i$}\\
1,&\mbox{otherwise.}
\end{cases}
\end{equation}
Let $\wtilde{\bm{\pi}}_k^j=(\wtilde{\pi}_{1,k}^j,\ldots,\wtilde{\pi}_{d,k}^j)$, $j=0,1$, be the constrained order-restricted MLE which maximizes \eqref{discLL} subject to the order-restriction $\pi_1\le\ldots\le\pi_d$ and the additional constraint that 
\begin{equation}\label{pi_i.cons}
\pi_{\what{i}^*_k}\le p_0\qmq{for} j=0\qmq{and}\pi_{\what{i}^*_k}\ge p_1\qmq{for}j=1,
\end{equation}
which can be computed as follows. If $\what{\pi}_{\what{i}^*_k,k}\le p_0$, then $\wtilde{\bm{\pi}}_k^0=\what{\bm{\pi}}_k$. Otherwise, $\what{\pi}_{\what{i}^*_k,k}> p_0$, so suppose that $\what{\pi}_{\what{i}^*_k-r-1,k}\le p_0<\what{\pi}_{\what{i}^*_k-r,k}$, in which case we set $\wtilde{\pi}_{\what{i}^*_k-r,k}^0=\ldots=\wtilde{\pi}_{\what{i}^*_k,k}^0=p_0$, and $\wtilde{\pi}_{i,k}^0$ coincides with $\what{\pi}_{i,k}$ for all other $i$. In other words, when $\what{\bm{\pi}}_k$ falls outside $H_0$, $\wtilde{\bm{\pi}}_k^0$ is computed by setting the appropriate elements of $\what{\bm{\pi}}_k$ to the boundary value~$p_0$, and $\wtilde{\bm{\pi}}_k^1$ is computed similarly.

The log-likelihood ratio statistics at the $k$th interim analysis for testing $H_0: \pi_{i^*}\le p_0$ vs.\ $H_1: \pi_{i^*}\ge p_1$ are given by 
\begin{equation}
\ell_{k,j}= \ell_k(\what{\bm{\pi}}_{k})- \ell_k(\wtilde{\bm{\pi}}_k^j),\quad j=0,1,
\end{equation} 
with $\ell_k(\bm{\pi})$ defined by \eqref{discLL}, and the group sequential test stops and rejects $H_0$ at interim analysis $k<K$ if
\begin{equation}\label{disc.b-rej}
\what{\pi}_{\what{i}_k^*,k}>p_0\qmq{and} \ell_{k,0}\ge b,
\end{equation} stops for futility if
\begin{equation}\label{disc.btild-rej}
\what{\pi}_{\what{i}_k^*,k}<p_1\qmq{and} \ell_{k,1}\ge \wtilde{b},
\end{equation} and otherwise rejects $H_0$ at the $K$th analysis if 
\begin{equation}\label{disc.c-rej}
\what{\pi}_{\what{i}_K^*,K}>p_0\qmq{and} \ell_{K,0}\ge c.
\end{equation} As in Section~\ref{sec:grp-sec-des}, the thresholds $b,\wtilde{b}$, and $c$ are chosen so that  \eqref{stage2-typeI} holds and  the power is close to $1-\beta$. Details are given in Section~\ref{sec:imp}.

 \subsection{Modeling the dependence between $y_i$ and $z_i$}\label{sec:disc.dep}
 A flexible method for modeling the general case where the toxicity and efficacy observations may not be independent is to introduce $d$ additional parameters in the form of the global cross ratios
 \begin{equation*}
 \rho_i=\frac{\Pi_i(0,0) \Pi_i(1,1)}{\Pi_i(1,0) \Pi_i(0,1)},\quad i=1,\ldots,d,\qmq{where} \Pi_i(y,z)=P(y_t=y, z_t=z|x_t=\lambda_i).
\end{equation*}  \citet{Dale86} proposed using the global cross ratio as a useful measurement of dependence in discrete ordered bivariate responses and they have been recently used by  \citet{Yin06} in a Bayesian Phase~I-II design. If the toxicity and efficacy responses are independent, then $\rho_i=1$ for all $i=1,\ldots,d$. The complete joint distribution $\Pi_i(y,z)$ of the toxicity and efficacy responses can be recovered from the parameters $\pi_i, \phi_i,\rho_i$, $i=1,\ldots,d$ through the following formulas:
\begin{gather*}
\Pi_i(1,1)=\begin{cases}
(a_i-\sqrt{a_i^2+b_i})/[2(\rho_i-1)],&\mbox{if $\rho_i\ne 1$}\\
\pi_i\phi_i,&\mbox{if $\rho_i= 1$}
\end{cases}\\
\Pi_i(1,0)=\pi_i-\Pi_i(1,1),\quad \Pi_i(0,1)=\phi_i-\Pi_i(1,1),\quad \Pi_i(0,0)=1-\pi_i-\phi_i+\Pi_i(1,1),
\end{gather*} where $a_i=1+(\pi_i+\phi_i)(\rho_i-1)$ and $b_i=-4\rho_i(\rho_i-1)\pi_i\phi_i$. The log-likelihood at the $k$th interim analysis of Phase~II for this general case is
\begin{equation}\label{discLL.dep}
\ell_k(\bm{\pi},\bm{\phi},\bm{\rho})=\log\left\{\prod_{t=1}^{\tau_k}\Pi_{i_t}(y_t,z_t)\right\}\qmq{where} x_t=\lambda_{i_t},
\end{equation} and the log-likelihood ratio statistics at the $k$th interim analysis for testing $H_0: \pi_{i^*}\le p_0$ vs.\ $H_1: \pi_{i^*}\ge p_1$ are given by 
\begin{equation}
\ell_{k,j}= \ell_k(\what{\bm{\pi}}_k,\what{\bm{\phi}}_k,\what{\bm{\rho}}_k)- \ell_k(\wtilde{\bm{\pi}}_k^j,\wtilde{\bm{\phi}}_k^j,\wtilde{\bm{\rho}}_k^j),\quad j=0,1,
\end{equation} with stopping rules as above in \eqref{disc.b-rej}-\eqref{disc.c-rej}, where $\what{\bm{\pi}}_k,\what{\bm{\phi}}_k,\what{\bm{\rho}}_k$ are MLEs maximizing \eqref{discLL.dep} subject to the order restrictions $\pi_1\le\ldots\le\pi_d$ and $\phi_1\le\ldots\le\phi_d$, and $\wtilde{\bm{\pi}}_k^j,\wtilde{\bm{\phi}}_k^j,\wtilde{\bm{\rho}}_k^j$ maximize \eqref{discLL.dep} subject to these order restrictions plus the constraints \eqref{pi_i.cons}.

\section{Simulation studies}\label{sec:sim}

\subsection{Operating characteristics of the traditional and proposed Phase~I-II designs on a  continuous dose space}\label{sec:sim_curr}

\label{p:E+S}To investigate the effect of uncertainty in the estimate $\what{\eta}$ on the operating characteristics of the Phase~II hypothesis test that is used in current practice, we first simulated a Phase~I design, which we take to be EWOC introduced by  \citet{Babb98}, followed by Simon's optimal 2-stage design. EWOC is a popular dose-finding method originally proposed for continuous dose spaces, which we consider here.  Motivated by a real trial for 5-flourouracil to treat solid colon tumors  described in  \citet{Babb98}, we let $[x_{\min},x_{\max}]=[140,425]$ denote the known range of acceptable dose values and assume $m=24$ patients are treated in Phase~I.  We parametrize the toxicity responses' distribution~$F(x;\cdot)$ by $\eta$ and $\rho=F(x_{\min};\bm{\theta})$ rather than $\bm{\theta}=(\theta_1,\theta_2)$ and assume that $(\rho,\eta)$ has the uniform distribution on $[0,q]\times [x_{\min},x_{\max}]$ as its prior distribution; see \cite[Section~2]{Bartroff10b} for more details. Fixing $\eta=250$, $q=1/3$ and $\rho=.1$, Table~\ref{table:E+S} gives some operating characteristics of this Phase~I-II design using Simon's design for testing \eqref{SimonHyp} with $p_0=.1$ and $p_1=.25$, with $\beta=.2$ and various values of $\alpha$. These were evaluated from 100,000 simulations using the above values of $\eta$, $x_{\min}$, and $x_{\max}$, and under the efficacy parameter $\bm{\psi}=(-3.895,.00679)$ chosen so that $p(\eta;\bm{\psi})=p_0=.1$ and $p(x_{\max};\bm{\psi})=.9$.

 \begin{table}\caption{Operating characteristics, based on 100,000 simulations, of the traditional design described on page~\pageref{p:E+S} in which $\eta=250$, $\what{\eta}$ denotes the final MTD estimate by either MLE, posterior mean (CRM), or EWOC, and $\alpha$ denotes the prescribed type~I error probability of Simon's Phase~II test of $p(\what{\eta};\bm{\psi})\le p_0$ with design parameters $n_1, n_2, r_1$ and $r$. The actual probability of rejecting $H_0: p(\eta;\bm{\psi})\le p_0$ is denoted by $P(-H_0|\what{\eta})$, with standard errors in parentheses, where $\what{\eta}=$ MLE, CRM, or EWOC is the final MTD estimate in Phase~I.}\label{table:E+S}
 \begin{center}
\begin{tabular}{ccccc}\hline
\multicolumn{2}{l}{Method for $\what{\eta}$}&MLE&CRM&EWOC\\\hline
\multicolumn{2}{l}{$\min(\what{\eta})$}&140.0&141.2&141.0\\
\multicolumn{2}{l}{$Q_1(\what{\eta})$}&226.3&246.9&229.1\\
\multicolumn{2}{l}{$\mbox{med}(\what{\eta})$}&244.7&264.7&246.9\\
\multicolumn{2}{l}{$Q_3(\what{\eta})$}&264.1&318.1&246.9\\
\multicolumn{2}{l}{$\max(\what{\eta})$}&425.0&391.6&362.7\\
\multicolumn{2}{l}{$E(\what{\eta})$}&252.6&276.7&239.8\\
\multicolumn{2}{l}{$\mbox{RMSE}(\what{\eta})$}&52.2&44.2&29.0\\\hline
$\alpha$&$n_1/n_2/r_1/r$&$P(-H_0|\mbox{MLE})$ &$P(-H_0|\mbox{CRM})$ &$P(-H_0|\mbox{EWOC})$\\\hline
.05&18/25/2/7&.180 (.001)&.479 (.002)&.100 (.0009)\\
.04&18/30/2/8&.176 (.001)&.476 (.002)&.094 (.0009)\\
.03&18/35/2/9&.170 (.001)&.470 (.002)&.088 (.0009)\\
.02&22/44/3/11&.167 (.001)&.464 (.002)&.083 (.0009)\\
.01&22/58/3/14&.156 (.001)&.458 (.002)&.074 (.0008)\\\hline
\end{tabular}
\end{center}
\end{table}

For several values of the parameters $(n_1,n_2,r_1,r)$ of Simon's two-stage design \cite[Table~2]{Simon89} of the Phase~II trial, Table~\ref{table:E+S} compares the prescribed type~I error probability~$\alpha$ of Simon's test with the actual probability of rejecting $H_0: p(\eta;\bm{\psi})\le p_0$,  denoted by $P(-H_0|\cdot)$,  for three choices of the MTD estimate $\what{\eta}$ that is used as the dose for the Phase~II trial. The three types of estimation are the MLE, the final posterior mean of the Phase~I trial (which is what the original version of the Bayesian CRM \citep{OQuigley90} would use), and the dose recommended by   EWOC that is used in the Phase~I design of this simulation study. Table~\ref{table:E+S} shows that the actual probability~$P(-H_0|\cdot)$ of falsely rejecting $H_0$ is largely inflated over the prescribed value~$\alpha$ of the type~I error probability used for the Phase~II trial, especially when the posterior mean is used for the MTD estimate $\what{\eta}$.  The reason for this is the frequent over-estimation of $\eta$ by $\what{\eta}$, as shown by the 5-number summary (maximum, first quartile~$Q_1$, median, third quartile $Q_3$, and maximum) of the 100,000 simulated values of $\what{\eta}$ given in the table. Although under-estimation of $\eta$ by $\what{\eta}$ also occurs, it is more often over-estimated, which causes rejection of $H_0$ at rates higher than prescribed by the design parameters of Simon's test. Also given in the table are the mean $E(\what{\eta})$ and the root-mean-square-error RMSE$(\what{\eta})=\{E(\what{\eta}-\eta)^2\}^{1/2}$ of the estimated MTD. We comment that here we have only considered the most basic versions of CRM \citet{OQuigley90} and EWOC \citet{Babb98}, and many variants have been proposed since then (e.g., \citet{Goodman95,Tighiouart10}). It seems likely that the properties of $\what{\eta}$ could be improved using one of these variants of CRM or EWOC, but since our focus here is more on the interaction between Phase~I and Phase~II, we do not explore that option here.

\begin{table}\caption{Operating characteristics of the traditional (denoted Trad) and new (denoted New) designs described on page~\pageref{p:BLNvES}.  The toxicity parameter is fixed at $(\eta,\rho)=(250,.1)$, and the six values of the efficacy parameter $\bm{\psi}$ are determined by $p(x_{\max};\bm{\psi})=.9$ and $p(\eta;\bm{\psi})=.05, .1, .2, .3, .4, .5$. All designs have Phase~I sample size $m=24$ and maximum Phase~II sample size $43$, for a maximum Phase~I-II sample size of $67$. Eff is the overall response rate for subjects in the study, OD is the overall overdose rate of subjects treated at doses above the true MTD, and RMSE$(\what{\eta}_{rec})$ is the root-mean-square-error of the recommended dose.}\label{table:BLNvES}
 \begin{center}
\begin{tabular}{lcccccccccccc}\hline
$p(\eta;\bm{\psi})$& \multicolumn{2}{c}{5\%}& \multicolumn{2}{c}{10\%}& \multicolumn{2}{c}{20\%}& \multicolumn{2}{c}{30\%}& \multicolumn{2}{c}{40\%}& \multicolumn{2}{c}{50\%}\\
&Trad&New&Trad&New&Trad&New&Trad&New&Trad&New&Trad&New\\\hline
$p(\what{\eta}_{rec};\bm{\psi})$&.101&.054&.150&.102&.233&.202&.319&.296&.409&.392&.499&.486\\
Eff&.096&.061&.140&.104&.219&.200&.310&.293&.405&.381&.498&.474\\
OD&.303&.291&.314&.312&.326&.289&.327&.256&.336&.252&.331&.249\\
RMSE$(\what{\eta}_{rec})$&51.0&28.4&52.2&29.0&52.4&29.3&52.3&28.6&51.7&29.0&52.1&29.8\\
$P(\mbox{rej.\ $H_0$})$&.090&.051&.180&.180&.479&.645&.776&.923&.939&.989&.987&.999\\
$EN$&45.9&40.2&49.8&47.3&57.7&51.0&63.2&43.7&66.0&37.0&66.7&34.6\\\hline
\end{tabular}
\end{center}
\end{table}

\label{p:BLNvES} Focusing on the traditional two-stage design with $\alpha=.05$ in Table~\ref{table:E+S} (denoted here by Trad) and concentrating on MLE estimation for simplicity, we compare its operating characteristics with those of the new Phase I-II design described in Section~\ref{sec:new-approach} (denoted by New).
 In order to match the Trad design's probability $P(\mbox{rej.\ $H_0$})=.18$ of falsely rejecting $H_0: p(\eta;\bm{\psi})\le p_0$, where $p_0=.1$, at the parameter values determined by $p(\eta;\bm{\psi})=.1$, we choose critical values $b=3$, $\wtilde{b}=3.5$, and $c=.7$ in \eqref{b-rej}-\eqref{c-rej}. Although a type~I error probability of .18 is usually deemed too high, we can keep the probability of falsely rejecting $H_0$ close to .05 if we use $p_0=.05$ instead, as shown in Table~\ref{table:BLNvES} which compares the operating characteristics of the Trad and New designs based on 10,000 simulations. The two designs both have Phase~I sample size of $m=24$ and maximum Phase~II sample size of 43, and the New design achieves this through Phase~II group sizes 10, 10, 10, 10, and 3. As in Table~\ref{table:E+S}, $\eta$ is fixed at 250 and $\rho=.1$, while $\bm{\psi}$ is specified by fixing $p(x_{\max},\bm{\psi})=.9$ and varying $p(\eta;\bm{\psi})$ over the values .05, .1, .2, .3, .4, and .5. For each scenario, Table~\ref{table:BLNvES} gives $P(\mbox{rej.\ $H_0$})$, in which $p_0=.1$, and the total expected sample size~$EN$ over the two phases. It shows that the new design has smaller $P(\mbox{rej.\ $H_0$})$ than Trad for $p(\eta;\bm{\psi})=.05$ and larger $P(\mbox{rej.\ $H_0$})$ for all values $p(\eta;\bm{\psi})>.1$, and uniformly smaller expected sample size, substantially so for parameter values $p(\eta;\bm{\psi})>.3$. In addition,  Table~\ref{table:BLNvES} also gives  the probability $p(\what{\eta}_{rec};\bm{\psi})$ of efficacious response at the recommended dose~$\what{\eta}_{rec}$ which for Trad is the MTD estimate at the end of Phase~I and for New is the final MLE at the end of Phase~II, the overall response rate (denoted Eff) for subjects in the study, the overall overdose rate (denoted OD) of subjects treated at doses above the true MTD, and the root-mean-square-error RMSE$(\what{\eta}_{rec})$ of the recommended dose.  The RMSE of the recommended dose for New is substantially smaller than Trad throughout, which we attribute to its continued estimation of $\eta$ during Phase~II. The values $p(\what{\eta}_{rec};\bm{\psi})$ and Eff are comparable to $p(\eta;\bm{\psi})$ throughout for New, while the corresponding values for Trad are larger, and  Trad has larger OD values than New.
 
Table~\ref{table:BLNvES}  shows a dramatic improvement of the New design relative to the Trad design in terms of both power and average sample size.  In order to discern how much of this improvement is due to the group sequential sampling used (relative to Simon's 2-stage design) versus how much is due to the continued estimation of the MTD during Phase~II that the proposed design allows, more simulation studies were performed whose results are in Tables~\ref{table:GRPvGRP} and \ref{table:SvS}.  In addition, both of these simulation studies were performed under different parameter values than in Table~\ref{table:BLNvES} in order to see the proposed design's performance over a broad range of scenarios.

\begin{table}\caption{Operating characteristics of the traditional (denoted Trad) and new (denoted New) designs described on page~\pageref{p:GRPvGRP}.  The toxicity parameter is fixed at $(\eta,\rho)=(350,.2)$, and the six values of the efficacy parameter $\bm{\psi}$ are determined by $p(x_{\max};\bm{\psi})=.95$ and $p(\eta;\bm{\psi})=.4, .5, .6, .7, .8, .9$. All designs have Phase~I sample size $m=24$ and maximum Phase~II sample size $43$, for a maximum Phase~I-II sample size of $67$. Eff is the overall response rate for subjects in the study, OD is the overall overdose rate of subjects treated at doses above the true MTD, and RMSE$(\what{\eta}_{rec})$ is the root-mean-square-error of the recommended dose. }\label{table:GRPvGRP}
 \begin{center}
\begin{tabular}{lcccccccccccc}\hline
$p(\eta;\bm{\psi})$& \multicolumn{2}{c}{40\%}& \multicolumn{2}{c}{50\%}& \multicolumn{2}{c}{60\%}& \multicolumn{2}{c}{70\%}& \multicolumn{2}{c}{80\%}& \multicolumn{2}{c}{90\%}\\
&Trad&New&Trad&New&Trad&New&Trad&New&Trad&New&Trad&New\\\hline
$p(\what{\eta}_{rec};\bm{\psi})$&.183&.179&.207&.219&.250&.270&.324&.361&.474&.511&.798&.812\\
Eff&.129&.123&.153&.151&.195&.195&.270&.280&.432&.439&.785&.788\\
OD&.111&.108&.109&.108&.111&.107&.109&.111&.109&.106&.102&.093\\
RMSE$(\what{\eta}_{rec})$&73.3&65.4&72.6&65.8&73.2&65.8&72.8&65.9&73.3&65.8&72.5&65.8\\
$P(\mbox{rej.\ $H_0$})$&.040&.043&.050&.050&.057&.062&.071&.088&.085&.134&.234&.590\\
$EN$&61.2&60.7&62.8&62.2&64.6&64.0&66.2&65.7&66.8&66.6&65.9&63.7\\\hline
\end{tabular}
\end{center}
\end{table}

\label{p:GRPvGRP} In Table~\ref{table:GRPvGRP}, the traditional Phase~I-II design (denoted Trad) was implemented but, instead of using Simon's two-stage design for Phase~II, the same group sequential sampling scheme that New used in Table~\ref{table:BLNvES} with group sizes 10, 10, 10, 10, and 3 was used. The proposed design (denoted New) was also implemented using these groups sizes and compared with Trad, so that the only difference between the two designs is that Trad does not update the estimate~$\what{\eta}$ of the MTD during Phase~II. To see the performance of the proposed design in a different scenario than Table~\ref{table:BLNvES}, using the same dose range $[x_{\min},x_{\max}]=[140,425]$ and prior structure as there, the true MTD $\eta$ was taken to be 350 and the probability of toxicity~$\rho$ at dose $x_{\min}$ was taken to be .2.  This scenario represents a much ``flatter'' dose-toxicity curve than in Table~\ref{table:BLNvES}.  In this set-up, the Phase~II null hypothesis $H_0:  p(\eta;\bm{\psi})\le p_0$ was tested with $p_0=.5$ and Table~\ref{table:GRPvGRP} contains the operating characteristics of these designs at six different values of the response parameter~$\bm{\psi}$ determined by $p(x_{\max},\bm{\psi})=.95$ and $p(\eta,\bm{\psi})=.4, .5, .6, .7, .8$ and $.9$. Unlike the Trad design in Table~\ref{table:BLNvES} which does not achieve the overall type~I error probability $P(\mbox{rej.\ $H_0$})$ at $p(\eta;\bm{\psi})=p_0$ equal to the prescribed value~$\alpha=.05$ because of the variance of the MTD estimate used in Phase~II, here the Trad design uses the stopping rule \eqref{b-rej}-\eqref{c-rej} with the critical values $b, \wtilde{b}$, and $c$ chosen so that this quantity is equal to $\alpha$ for $p_0=.5$; they are $b=24.1$, $\wtilde{b}=64.4$, and $c=21.8$.  The New design uses the values $b=18.8$, $\wtilde{b}=54.4$, and $c=11.7$, also chosen so that its type~I error probability is $\alpha$, and are slightly different than Trad's critical values because New continues to update $\what{\eta}$ during Phase~II.  Table~\ref{table:GRPvGRP} contains the operating characteristics of these designs based on 10,000 Monte Carlo replications at each parameter value.  As might be expected from designs using the same sampling scheme, Trad and New have very similar expected sample size, and sample sizes are in general larger in this scenario than the one in Table~\ref{table:BLNvES} which is also to be expected because of the flatness of the dose-toxicity curve which makes $\eta$ difficult to estimate accurately, reflected in the power of both designs being low until $p(\eta,\bm{\psi})$ reaches 90\%, where the power of New is 59\% but Trad is still severely underpowered. Note also that even though the flatness of the dose-toxicity makes the MTD difficult to estimate accurately, the chance of overdose is relatively low. Overall, New is slightly but consistently more efficient with smaller RMSE despite having slightly smaller average sample size, and New has higher power and response probabilities~$p(\what{\eta}_{rec};\bm{\psi})$  over the range of parameter values in the alternative.  These results are consistent with the two designs using the same Phase~II sampling scheme but New using continued estimation of the MTD throughout Phase~II.
 
 \begin{table}\caption{Operating characteristics of the traditional (denoted Trad) and new (denoted New) designs described  on page~\pageref{p:SvS}. The toxicity parameter is fixed at $(\eta,\rho)=(200,.15)$, and the six values of the efficacy parameter $\bm{\psi}$ are determined by $p(x_{\max};\bm{\psi})=.99$ and $p(\eta;\bm{\psi})=.025, .05, .25, .45, .65, .85$. All designs have Phase~I sample size $m=24$ and maximum Phase~II sample size $30$, for a maximum Phase~I-II sample size of $54$. Eff is the overall response rate for subjects in the study, OD is the overall overdose rate of subjects treated at doses above the true MTD, and RMSE$(\what{\eta}_{rec})$ is the root-mean-square-error of the recommended dose. }\label{table:SvS}
 \begin{center}
\begin{tabular}{lcccccccccccc}\hline
$p(\eta;\bm{\psi})$& \multicolumn{2}{c}{2.5\%}& \multicolumn{2}{c}{5\%}& \multicolumn{2}{c}{25\%}& \multicolumn{2}{c}{45\%}& \multicolumn{2}{c}{65\%}& \multicolumn{2}{c}{85\%}\\
&Trad&New&Trad&New&Trad&New&Trad&New&Trad&New&Trad&New\\\hline
$p(\what{\eta}_{rec};\bm{\psi})$&.011&.032&.017&.061&.082&.263&.197&.454&.383&.648&.706&.849\\
Eff&.008&.038&.011&.067&.061&.268&.178&.458&.377&.656&.706&.850\\
OD&.612&.611&.613&.615&.367&.564&.595&.569&.575&.578&.570&.570\\
RMSE$(\what{\eta}_{rec})$&31.3&25.8&31.7&27.6&30.1&34.7&30.8&35.4&30.9&33.7&30.7&35.3\\
$P(\mbox{rej.\ $H_0$})$&.007&.002&.010&.010&.209&.185&.636&.478&.940&.763&.999&.819\\
$EN$&33.9&34.1&34.8&36.8&42.7&52.8&49.4&54.0&53.2&53.9&53.9&53.9\\\hline
\end{tabular}
\end{center}
\end{table}

\label{p:SvS} Table~\ref{table:SvS} considers another scenario, with a smaller Phase~II sample size, in which both Trad and New use for  Phase~II a two-stage design with early stopping only for futility. For Trad this is Simon's two-stage design and for New this is the stopping rule \eqref{b-rej}-\eqref{c-rej} with $K=2$ groups and $b$ fixed at $\infty$ so that only early stopping for futility can occur. In this scenario both Trad and New have maximum Phase~II sample size 30 (compared with 43 in Tables~\ref{table:BLNvES} and \ref{table:GRPvGRP}) and Phase~I sample size $m=24$. To achieve this, Trad uses Simon's \cite[Table~1]{Simon89} design with $r_1=0$, $n_1=9$, $r=3$, and $n_2=21$ for $\alpha=.05$ and $\beta=.1$ at $p_0=.05$ and $p_1=.25$. As in Tables~\ref{table:E+S} and \ref{table:BLNvES}, the Trad design using these parameters does not achieve the type~I error probability at the prescribed value $\alpha=.05$ because of variance of the MTD estimate used in Phase~II.  Indeed, Table~\ref{table:SvS} shows its actual type~I error probability to be .01 at $p(\eta;\bm{\psi})=p_0=.05$. Unlike Table~\ref{table:BLNvES} that shows inflation of type~I error probability, here the type~I error probability is substantially smaller then the prescribed value $\alpha=.05$. In order to make a meaningful  comparison between designs we choose the parameters of the New design to match this smaller value of the type~I error probability, for which we use $b=\infty$ (to allow early stopping only for futility), $\wtilde{b}=2.1$ and $c=19.3$ in \eqref{b-rej}-\eqref{c-rej} and Phase~II group sizes 9 and 21, the same as the Simon design.  The operating characteristics of these designs are given in Table~\ref{table:SvS}, based on 10,000 Monte Carlo replications each, in yet another scenario with  $\eta=200$, $\rho=.15$, and six values of $\bm{\psi}$ determined by $p(x_{\max};\bm{\psi})=.99$ and $p(\eta;\bm{\psi})=.025, .05, .25, .45, .65$, and $.85$.  The dose range and prior structure are the same as in Table~\ref{table:GRPvGRP}. The response probabilities~$p(\what{\eta}_{rec};\bm{\psi})$ at New's recommended dose  stay much closer to the true values than at Trad's recommended dose, likely due to New's update of the MTD estimate during Phase~II. The overall response rate of subjects in the study is also substantially higher in New than in Trad. The two designs have similar average sample sizes, reflective of their similar sampling schemes, and Trad has higher power in the alternative. The RMSEs of the two designs are small and relatively close, with Trad's being slightly smaller.  Note, however, that the squared error RMSE$(\what{\eta}_{rec})$ ignores the sign of $\eta-\what{\eta}_{rec}$ and that the results on $p(\what{\eta}_{rec};\bm{\psi})$ show that $\what{\eta}_{rec}$ tends to under-estimate $\eta$.

\subsection{Performance of the traditional and new Phase~I-II designs on discrete dose space under monotonicity constraints}\label{sec:iso.sim} To evaluate the performance of the Phase~II method proposed in Section~\ref{sec:ordered} for monotonic efficacy and toxicity models on a discrete dose space, we performed a similar study to the one in Section~\ref{sec:sim_curr}, assuming independence of the toxicity and efficacy responses for simplicity; we have performed additional simulations under dependent responses using the model described in Section~\ref{sec:disc.dep} and the performance of the new method is similar. Again focusing on the Trad design in Table~\ref{table:E+S} and using isotonic MLE estimation~\eqref{iso.MLE} for both the Trad and new (denoted by New) designs, the estimated operating characteristics are compared in Table~\ref{table:iso} based on 10,000 simulated trials, wherein the Phase~I doses of the $m=24$ patients are uniformly sampled from the dose set $\Lambda=\{140,200,250,300,350,425\}$. In this setting, the Trad design with nominal level $\alpha=.05$ for testing $\pi_{\what{i}^*}\le p_0$ actually has type~I error probability $P(\mbox{rej.\ $H_0$})=.211$ of falsely rejecting $H_0: \pi_{i^*}\le p_0=.1$, and so in order to compare New and Trad in this setting we choose critical values $b=.13$, $\wtilde{b}=3.3$, and $c=.03$ in \eqref{b-rej}-\eqref{c-rej} in order to approximately match this, giving $P(\mbox{rej.\ $H_0$})=.201$ for New at $\pi_{i^*}=.1$. In order to have the same maximum Phase~II sample size~$M=43$ as Trad, again New uses group sequential sampling with group sizes 10, 10, 10, 10, and 3. In this discrete nonparametric setting, the unknown parameters are the true toxicity and efficacy probabilities $\bm{\phi}$ and $\bm{\pi}$ given by \eqref{disc.nonparm}, and in order to compare Trad and New in a setting similar to the one in Section~\ref{sec:sim_curr}, we consider values of $\bm{\phi}$ and $\bm{\pi}$ given by the corresponding parametric models $F(x;\bm{\theta})$ and $p(x;\bm{\psi})$ and parameter values given there: $\eta=\lambda_{i^*}$ is fixed at 250, $\rho=\phi_1=.1$, $p(x_{\max},\bm{\psi})=\pi_d=.9$, and $p(\eta;\bm{\psi})=\pi_{i^*}=.05, .1, .2, .3, .4$, and .5.  The relative performance of Trad and New is very similar to that in the previous section: The new design has smaller $P(\mbox{rej.\ $H_0$})$ than Trad for parameter values $\pi_{i^*}\le .1$ in the null hypothesis, larger $P(\mbox{rej.\ $H_0$})$ for all values $\pi_{i^*}> .1$ in the alternative, and uniformly smaller expected sample size, substantially so when $\pi_{i^*}$ is large or small relative to $p_0=.1$. The other operating characteristics given in the table are the same as in Table~\ref{table:BLNvES}: The response rate~$\pi_{\what{i}^*}$ at the final recommended dose, overall response rate (Eff) and overdose rate (OD) of patients in the study, and the RMSE of the final recommended dose~$\lambda_{\what{i}^*}$. The Eff rate of New is larger than Trad at all parameter values considered, which we attribute to the proposed design's ability to vary the dose throughout Phase~II, and hence ``correct'' for a poorly chosen MTD estimate at the end of Phase~I, to some measure. The OD rates of the two designs are close, with New being sometimes smaller and sometimes larger. The RMSE of New is slightly larger, but comparable to Trad, which we attribute to its markedly smaller average sample size.

\begin{table}\caption{Operating characteristics of the traditional (denoted Trad) and new (denoted New) designs described in Section~\ref{sec:iso.sim}.  The true toxicity and efficacy probabilities are determined by the same parameters as in Table~\ref{table:BLNvES}: $\lambda_{i^*}=250$, $\phi_1=.1$, $\pi_d=.9$, and  the six cases $\pi_{i^*}=.05, .1, .2, .3, .4$, and $.5$ as described in Section~\ref{sec:iso.sim}. Eff is the overall response rate for subjects in the study, OD is the overall overdose rate of subjects treated at doses above the true MTD, and RMSE$(\what{\eta}_{rec})$ is the root-mean-square-error of the recommended dose. }\label{table:iso}
 \begin{center}
\begin{tabular}{lcccccccccccc}\hline
$\pi_{i^*}$& \multicolumn{2}{c}{5\%}& \multicolumn{2}{c}{10\%}& \multicolumn{2}{c}{20\%}& \multicolumn{2}{c}{30\%}& \multicolumn{2}{c}{40\%}& \multicolumn{2}{c}{50\%}\\
&Trad&New&Trad&New&Trad&New&Trad&New&Trad&New&Trad&New\\\hline
$\pi_{\what{i}^*}$&.072&.030&.116&.061&.194&.131&.274&.206&.357&.295&.449&.395\\
Eff&.185&.196&.225&.231&.286&.296&.350&.364&.416&.441&.492&.524\\
OD&.390&.376&.388&.363&.366&.361&.347&.370&.328&.390&.320&.406\\
RMSE$(\lambda_{\what{i}^*})$&56.5&60.1&57.4&60.9&56.7&59.1&56.9&59.2&56.5&58.3&57.2&57.9\\
$P(\mbox{rej.\ $H_0$})$&.117&.076&.211&.201&.410&.486&.615&.729&.805&.895&.931&.981\\
$EN$&56.5&38.7&49.4&40.7&54.8&41.7&59.8&40.2&63.6&37.7&65.9&35.6\\\hline
\end{tabular}
\end{center}
\end{table}

\section{Group sequential likelihood theory and implementation details}\label{sec:imp+theory}

\subsection{Theory of group sequential GLR tests}\label{sec:theory}
We first assume independence between $y_i$ and $z_i$ given $x_i$ as in Section~\ref{sec:grp-sec-des}. In this case, the likelihood function, based on a sample of size $\tau_k$, is of the form $L_{1,k}(\bm{\theta}) L_{2,k}(\bm{\psi})$, where
\begin{equation*}
L_{1,k}(\bm{\theta})=\prod_{i=1}^{\tau_k} [F(x_i;\bm{\theta})]^{y_i} [1-F(x_i;\bm{\theta})]^{1-y_i},\quad  L_{2,k}(\bm{\psi})=\prod_{i=1}^{\tau_k} [p(x_i;\bm{\psi})]^{z_i} [1-p(x_i;\bm{\psi})]^{1-z_i}.
\end{equation*} The GLR statistic for testing $p(\eta;\bm{\psi})=p_j$, which is the boundary of $H_j$, is
\begin{equation}\label{LGLR}
\log\left[\left.\left\{\sup_{\bm{\theta}} L_{1,k}(\bm{\theta})\times \sup_{\bm{\psi}} L_{2,k}(\bm{\psi})\right\}\right/ \left\{\sup_{(\bm{\theta},\bm{\psi}):\; p(\eta;\bm{\psi})=p_j} L_{1,k}(\bm{\theta}) L_{2,k}(\bm{\psi})\right\}\right],
\end{equation}
and the signed root likelihood ratio statistic is approximately normal under $p(\eta;\bm{\psi})=p_j$; see \cite[p.~513]{Lai04}. Note that $p(\eta;\bm{\psi})=p_j$ can be expressed as an equality constraint $\psi_1+\eta\psi_2=\mbox{logit}(p_j)$ on the linear function~$\psi_1+\eta\psi_2$ of $\bm{\psi}$, and we can reparameterize  $\bm{\psi}$ as $(\psi_1, \psi_1+\eta\psi_2)$ and $\bm{\theta}$ as $(\eta,\rho)$. Therefore, standard asymptotic analysis of GLR statistics shows that under $p(\eta;\bm{\psi})=p_j$, \eqref{LGLR} has the same limiting distribution as
\begin{equation}\label{LGLR'}
\log\left[\left. \sup_{\bm{\psi}} L_{2,k}(\bm{\psi}) \right/ \sup_{\bm{\psi}:\; p(\what{\eta}_k;\bm{\psi})=p_j} L_{2,k}(\bm{\psi}) \right],
\end{equation} jointly over $1\le k\le K$; see \cite[Section~9.3(iii)]{Cox74}. Because the $x_i$ are sequentially determined random variables (based on group sequential estimates of the MTD), we use the martingale central limit theorem \citep[][p.~411]{Durrett05},  here instead of the traditional central limit theorem as in \citet{Lai04}. Note that \eqref{LGLR'} is the same as $\ell_{k,j}$ defined in \eqref{GLRs}. For the dependent case in Section~\ref{sec:dependent}, the likelihood function $L_{2,k}(\bm{\psi})$ involves both $z_i$ and $y_i$ in view of \eqref{z|y=0} and \eqref{z|y=1} but does not depend on $\eta$. A similar argument can be used to show that the GLR statistic at the $k$th interim analysis is still asymptotically equivalent to \eqref{LGLR'}.

The group sequential GLR test of $H_0$ is much more flexible and efficient than Simon's 2-stage likelihood ratio test~\citep{Simon89} for Phase~II cancer trials. As noted in the last paragraph of Section~\ref{sec:integrated}, Simon's procedure actually tests $p(\what{\eta};\bm{\psi})\le p_0$ with all doses set at the MTD estimate $\what{\eta}$ from the Phase~I toxicity data, while the proposed test considers the more natural $H_0: p(\eta;\bm{\psi})\le p_0$ and uses all the observed $(x_i,y_i,z_i)$ up to the time of interim analysis to test $H_0$. Moreover, unlike Simon's two-stage design which is actually a group sequential test with two groups and only allows futility stopping in the first stage, we use a more flexible group sequential design that allows early stopping for both efficacy and futility. In addition, the estimate of $\eta$ of the Phase~I-II trial uses data up to the end of the trial. The group sequential GLR test uses the alternative $p_1$ implied by the maximum size $\tau_K$ (see Section~\ref{sec:imp}) to derive the futility stopping criterion, namely stopping when there is enough evidence against $H_1: p(\eta;\bm{\psi})\ge p_1$. Similarly, it stops early for efficacy if the GLR statistics show enough evidence against $H_0: p(\eta;\bm{\psi})\le p_0$.

The group sequential GLR test in Section~\ref{sec:ordered} that considers discrete dose levels also involves a finite number of parameters satisfying certain monotonicity constraints. Therefore the theory of group sequential tests that we have applied to the logistic regression models in Section~\ref{sec:new-approach} can also be applied to Section~\ref{sec:ordered} that imposes certain structure on the parameter space. Lai and Shih~\cite[Section~3]{Lai04} have established the asymptotic efficiency of these group sequential GLR tests in terms of the expected sample size and power function. Here we extend this theory in two ways. The first extension is from the i.i.d.\ model to the regression model, with sequentially determined regressors $x_i$. The second extension is to replace the GLR statistics by more easily computable and interpretable approximations that have the same asymptotic distributions. As noted above, martingale theory used in conjunction with likelihood theory provides the key tools for such extensions.

\subsection{Implementation details}\label{sec:imp}
The MLEs of $\bm{\theta}$ and $\bm{\psi}$ involved  in the design proposed in Section~\ref{sec:new-approach} should be computed under the assumption of positive slope, i.e., $\theta_2>0$ and $\psi_2>0$. In practice this can be imposed by choosing a small value $\delta>0$ and computing the MLEs under the constraint $\theta_2\ge\delta$ and $\psi_2\ge\delta$.  A related issue is that the MLEs of $\bm{\theta}$ and $\bm{\psi}$ may not exist in the first few stages of Phase~I \citep[see p.~195 of][]{Agresti02}.  In this case, their Bayes estimates from a Bayesian model-based design can be used instead.

The alternative $p_1>p_0$ is that implied by the maximum sample size~$\tau_K$ and the desired type~I and II error probabilities $\alpha$ and $\beta$, respectively. That is, for the GLR test that has fixed sample size $\tau_K$ and  rejects $H_0$ if and only if 
\begin{equation}\label{FSS}
p(\what{\eta}_K;\what{\bm{\psi}}_K)>p_0\qmq{and}\min_{ \bm{\psi}\in\mathcal{S}_K^0}\left[ \ell_K(\what{\bm{\psi}}_{K})-\ell_K(\bm{\psi})\right] \ge C_\alpha,
\end{equation} let $p_1>p_0$ be the alternative satisfying
\begin{equation}\label{FSS-power}
\min_{\bm{\psi}\in\mathcal{S}_0^1} P_{\wtilde{\bm{\theta}},\bm{\psi}}[\mbox{(\ref{FSS}) occurs}|\mathcal{F}_0]=1-\beta.
\end{equation}
In \eqref{FSS}, $C_\alpha$ is such that  
\begin{equation}\label{FSS-typeI}
\max_{\bm{\psi}\in\mathcal{S}_0^0} P_{\wtilde{\bm{\theta}},\bm{\psi}}[\mbox{(\ref{FSS}) occurs}|\mathcal{F}_0]= \alpha\end{equation} and the doses $x_1,\ldots,x_{\tau_K}$ are chosen by some design.  The computation of the left-hand sides of \eqref{FSS-power} and \eqref{FSS-typeI} will be described below.

The thresholds $b,\wtilde{b}$, and $c$ in \eqref{b-rej}-\eqref{c-rej} can be determined as follows. Let $0<\eps<1/2$ and  first choose $\wtilde{b}$ so that 
\begin{equation}\label{rejH0early}
\max_{\bm{\psi}\in\mathcal{S}_0^1} P_{\wtilde{\bm{\theta}},\bm{\psi}}[\mbox{(\ref{btild-rej}) occurs for some $1\le k<K$}|\mathcal{F}_0]= \eps\beta.
\end{equation} Then choose $b$ so that 
\begin{equation}
\max_{\bm{\psi}\in\mathcal{S}_0^0} P_{\wtilde{\bm{\theta}},\bm{\psi}}[\mbox{(\ref{b-rej}) occurs for some $1\le k<K$, $p(\what{\eta}_{\tau_{k'}},\what{\bm{\psi}}_{\tau_{k'}})\ge p_1$ and $\ell_{k',1}< \wtilde{b}$ for all $k'<k$}|\mathcal{F}_0]= \eps\alpha,
\end{equation} and finally choose $c$ so that
\begin{equation}\label{c-prob}
\max_{\bm{\psi}\in\mathcal{S}_0^0} P_{\wtilde{\bm{\theta}},\bm{\psi}}[\mbox{(\ref{c-rej}) occurs and (\ref{b-rej}), (\ref{btild-rej}) do not occur for any $1\le k <K$}|\mathcal{F}_0]=(1-\eps)\alpha.
\end{equation} The determination of $b$, $\wtilde{b}$ and $c$ in \eqref{rejH0early}-\eqref{c-prob} follows that in \citet{Lai04} and aims at controlling the type~I error probability~\eqref{stage2-typeI} and keeping the power~\eqref{stage2-pow} close to $1-\beta$.

As in Section~3.4 of \citet{Lai04}, we can use the joint asymptotic normality of the signed root likelihood ratio statistics to approximate the probabilities in \eqref{FSS-power}-\eqref{c-prob}. Because the GLR statistics are asymptotic pivots, the convergence in distribution holds uniformly over $\mS_0^1$ or $\mS_0^0$ and therefore the minimum (or maximum) over $\mS_0^1$ or $\mS_0^0$ in the left-hand sides of \eqref{FSS-power}-\eqref{c-prob} poses no additional difficulty when we use the normal approximation. An alternative to normal approximation is to use Monte Carlo similar to that used in the bootstrap tests.  Bootstrap theory suggests that we can simulate from the estimated distribution under an assumed composite hypothesis since the GLR statistic is an approximate pivot under that hypothesis. Thus, the bootstrap test chooses the $\bm{\psi}\in\mS_0^j$ in \eqref{FSS-power}-\eqref{c-prob} to be the MLE based on the Phase~I data $\mF_0$, of $\bm{\psi}$ under the constraint $p(\wtilde{\eta};\bm{\psi})=p_j$. In the simulation studies in Section~\ref{sec:sim}, we use 10,000 bootstrap simulations to estimate the probabilities in \eqref{FSS-power}-\eqref{c-prob}. The implementation of the group sequential order-restricted GLR test of $H_0: \pi_{i^*}\le p_0$ in Section~\ref{sec:ordered} is similar, as we have explicit formulas \eqref{discLL} and \eqref{ordMLE}. A software package to design the proposed Phase~I-II trial has been developed using \textsf{R} and is available at the website \url{http://med.stanford.edu/biostatistics/ClinicalTrialMethodology.html}.

\section{Discussion}\label{sec:disc}
The simulation studies in Section~\ref{sec:sim}, which are motivated by the trial in \citet{Babb98}, show that the estimate $\what{\eta}$ at the end of the Phase~I trial can substantially over- or under-estimate $\eta$ and therefore have a significantly higher or lower response rate than $p(\eta;\bm{\psi})$. Another situation in which the latter can occur is when using the  3+3 dose escalation scheme in Phase~I, which tends to produce a sub-therapeutic dose $\what{\eta}$ at the end of Phase~I.  Continuing dose-finding in Phase~II can add substantial information for estimating $\eta$, as Section~\ref{sec:sim} has shown.

Recognizing that the dose chosen at the end of the Phase~I trial may not ensure safety, \citet{Bryant95} have extended Simon's two-stage design for the Phase~II trial to incorporate toxicity outcomes in the Phase~II trial by stopping the trial after the first stage if either the observed response rate is inadequate or the number of observed toxicities is excessive, and by recommending the treatment at the end of the Phase~II trial only if there are both a sufficient number of responses and an acceptably small number of toxicities. Note that the Bryant-Day design still uses $\what{\eta}$ determined from the Phase~I data to be the dose throughout the Phase~II trial.  We have developed herein a novel methodology which continues dose finding to estimate the MTD in Phase~II and which uses the toxicity outcomes throughout the trial in a natural way, while focusing on testing the efficacy hypothesis during the Phase~II component of the Phase~I-II design. The methodology enables the user to carry out the novel group sequential extensions, allowing early stopping not only for futility but also for efficacy, of Simon's two-stage design  that is widely used in Phase~II cancer trials. These group sequential tests use efficient GLR statistics, which we have extended herein from the traditional logistic regression models in Section~\ref{sec:new-approach} to robust isotonic regression models in Section~\ref{sec:ordered}.

Bayesian designs have been proposed for Phase~II trials, allowing early stopping for efficacy or futility, and rejecting (or accepting) the hypothesis $p\le p_0$ if the posterior probability of $p>p_0$ exceeds some threshold (or falls below another threshold), thereby extending the Bayesian approach from Phase~I to Phase~II trials; see Chapter~4 of \citet{Berry10}.  \citet{Yin06} and \citet{Yin09b} have developed Bayesian Phase~I-II designs to incorporate the bivariate outcomes of toxicity and efficacy to determine the dose sequentially for the next cohort of patients in the trial.  Their underlying philosophy is that ``with a very limited sample size in the (traditional) phase~I trial, the MTD might not be obtained in a reliable way,'' and therefore they aim instead at finding ``the optimal dosage of a drug which has the highest effectiveness as well as tolerable toxicity'' \cite[p.~777]{Yin06}. Two motivating trials that attempt to ``speed up the drug discovery and reduce the total cost'' are given in \citet[][p.~925 and Section~3]{Yuan11} and \citet{Yin06}.

The trials that motivate the Phase~I-II design proposed herein are traditional Phase~I and Phase~II trials at cancer centers of most medical schools, such as the Norris Comprehensive Cancer Center at the University of Southern California and the Cancer Institute at Stanford University. The protocols usually have small sample sizes for Phase~I, followed by Simon's two-stage design for Phase~II that uses the MTD estimated from the Phase~I data. Simon's design has been popular because it allows interim analysis for a go/no go decision while preserving the type~I error probability and power at the effect size used to justify the sample size specified in the protocol.  The reason why investigators with whom we have worked adhere to this design although they recognize difficulties with the relatively small sample sizes for both phases is that they can publish the trial results in medical journals that prefer frequentist testing. The Phase~I-II design proposed herein is an attempt to enable the investigators to perform valid group sequential tests of efficacy while continuing estimation of the MTD during the entire course of the Phase~I-II trial.  Even though pharmaceutical companies do not need to publish the results of Phase~II trials and can focus on dose finding that incorporates both toxicity and efficacy as in the Bayesian designs of \citet{Yin06} and \citet{Yin09b}, many industry-sponsored Phase~II trials are still conducted at academic centers where this innovative Phase~I-II design can allow investigators to carry out group sequential frequentist testing of efficacy at the MTD and update the MTD estimate during the entire course of the trial.  While the present paper has established the basic methodology, much of the work for its adoption still lies ahead. This includes generating some experience in actual trials and their protocols, holding monthly forums and regular consulting sessions for clinical investigators at the U.S.C.\ Norris Cancer Center and the Stanford Cancer Institute, and developing user-friendly software based on this experience, which will facilitate its use by other academic centers.

\section*{Acknowledgements} Bartroff's work was supported by NSF grants DMS-0907241 and DMS-1310127 and
NIH grant GMS-068968.   Lai's work was supported by NSF grant DMS-1106535  and NIH grant 5P30CA124435. Narasimhan's work was supported by NCI Cancer Center Support Grant 5P30CA124435. 

%\bibliographystyle{apalike}
%\bibliography{../Bib_files/bibliography}

\def\cprime{$'$}

\end{document}